\documentclass{elsart}
\usepackage{epsfig}
\newcommand{\ph}{{\hat p}}
\begin{document}

\begin{frontmatter}

\title{Random Matrix Theory, Chiral Perturbation Theory, and Lattice
  Data}

\author[KL]{M.E.~Berbenni-Bitsch},
\author[R]{M.~G\"ockeler},
\author[R]{H.~Hehl},
\author[KL]{S.~Meyer},
\author[R]{P.E.L.~Rakow},
\author[R]{A.~Sch\"afer}, and
\author[M]{T.~Wettig}


\date{20 July 1999}

\address[KL]{Fachbereich Physik -- Theoretische Physik, Universit\"at
  Kaiserslautern, D-67663 Kaiserslautern, Germany}
\address[R]{Institut f\"ur Theoretische Physik, Universit\"at
  Regensburg, D-93040 Regensburg, Germany}
\address[M]{Institut f\"ur Theoretische Physik, Technische
  Universit\"at M\"unchen, D-85747 Garching, Germany}

\begin{abstract}
  Recently, the chiral logarithms predicted by quenched chiral
  perturbation theory have been extracted from lattice calculations of
  hadron masses. We argue that the deviations of lattice results from
  random matrix theory starting around the so-called Thouless energy
  can be understood in terms of chiral perturbation theory as well.
  Comparison of lattice data with chiral perturbation theory formulae
  allows us to compute the pion decay constant. We present results
  from a calculation for quenched SU(2) with Kogut--Susskind fermions
  at $\beta=2.0$ and $2.2$.

\end{abstract}

\end{frontmatter}
\newpage

For many observables, quenched chiral perturbation theory predicts
contributions which are logarithmic in the quark mass
\cite{Golt,Sharp}.  Their identification in lattice gauge results is a
long standing problem. It seems that the latest numerical results
\cite{Bard,Schier,Lat1,Lat2} on hadron masses in quenched lattice
simulations allow for an approximate determination of these $\ln(m)$
contributions. (For an earlier attempt see \cite{kim}.)  The
determination of these logarithms is an important test of chiral
perturbation theory (chPT) which in turn plays a central role for the
connection of low-energy hadron theory on one side and perturbative
and lattice QCD on the other.

In a completely independent development, it has been shown by several
authors that chiral random matrix theory (chRMT) is able to reproduce
quantitatively microscopic spectral properties of the Dirac operator
obtained from QCD lattice data (see the reviews~\cite{r1,r2} and
Refs.~\cite{r3,r4,r5,Ma}). Moreover, the limit up to which the
microscopic spectral correlations can be described by random matrix
theory (the analogue of the Thouless energy of statistical physics)
was analyzed theoretically in \cite{Zahed98,Jac98} and identified for
quenched SU(2) (SU(3)) lattice calculations in \cite{Ber2}
(\cite{su3pap}).  It has also been shown \cite{chiralpert} that the
results obtained in chRMT can be derived directly from field theory,
providing a firm theoretical basis for the RMT approach.

In this letter we want to study the Dirac spectrum beyond the Thouless
energy using chPT (for a first investigation of this issue see
\cite{chlog1}).  Lattice Monte Carlo calculations inevitably involve a
finite volume, so we have to consider chPT in a finite volume, too.
Chiral RMT should be valid for masses $m$ up to the Thouless energy
$\sim 1/L^2$ ($L$ = linear extent of the lattice), whereas the chPT
formulae are supposed to work if $m$ is larger than a certain lower
limit scaling like $1/L^4$. Thus we expect a domain of common
applicability for sufficiently large lattices.  The data we are going
to analyze were obtained for the gauge group SU(2) in the quenched
approximation with staggered fermions. They consist of complete
spectra of the lattice Dirac operator. As the bare coupling in these
data is relatively strong, we shall set up our (quenched) chPT in such
a way that only those symmetries are taken into account which are
exactly realized on the lattice. We will find that this eliminates
terms $\propto \ln (m)$ from the quenched chiral condensate, which are
present if one starts from the continuum symmetries.

In the following we shall use the chiral condensate and the scalar
susceptibilities, so we first give their definitions.  For a finite
lattice and a non-vanishing valence-quark mass, the chiral condensate
is given by
\begin{equation}
 \sigma_{\mathrm {lattice}}(m) = \frac{1}{N} \left\langle
 \sum_{k=1}^N \frac{1}{{\mathrm i}\lambda_k+m} \right\rangle \;.
\end{equation}
In the quenched theory, the connected susceptibility is given simply
by
\begin{equation}
 \chi^{\mathrm {conn}}_{\mathrm {lattice}}(m)
  = \frac{\partial}{\partial m} \sigma(m)
  = -\frac1N\left\langle\sum_{k=1}^N\frac1{({\mathrm i}\lambda_k+m)^2}
                    \right\rangle\;.
\end{equation}
The disconnected susceptibility is defined on
the lattice by
\begin{equation}
  \chi^{\mathrm {disc}}_{\mathrm {lattice}}
  =\frac{1}{N}\left\langle\sum_{k,l=1}^N
    \frac{1}{({\mathrm i}\lambda_k+m)({\mathrm i}\lambda_l+m)}
      \right\rangle -
  \frac{1}{N} \left\langle\sum_{k=1}^N\frac{1}{{\mathrm
        i}\lambda_k+m}\right\rangle^2 \;.
\label{e4}
\end{equation}
Here, $N=L^4$ denotes the number of lattice points, and the
$\lambda_k$ are the Dirac eigenvalues. Note that each of the doubly
degenerate eigenvalues (for gauge group SU(2)) counts only once.

With the abbreviation
$ F(x) = \int^1_0 \! \d t\: I_0(tx) $
the chRMT result for the chiral condensate reads
\begin{equation}
  \frac{\sigma_{\mathrm {RMT}}}{\Sigma} = 2u \Bigl[ I_0(2u) K_0(2u) +
    I_1(2u) K_1(2u) - K_0(2u) F(2u) \Bigr] \,,
\end{equation}
where $\Sigma$ denotes the absolute value of the chiral condensate for
infinite volume and vanishing mass, and the rescaled mass parameter
$u$ is given by $u=m\Sigma L^4$. The functions $I_\nu$, $K_\nu$ are
modified Bessel functions. For the connected susceptibility chRMT
predicts
\begin{equation}
  \frac{\chi^{\mathrm {conn}}_{\mathrm {RMT}}}{N\Sigma^2} =
  2 K_1 (2u) \Bigl[ 2u F(2u) - I_1 (2u) \Bigr] \,.
\end{equation}
The chRMT result for the disconnected susceptibility is slightly more
complicated \cite{Ber2}, but can be simplified to read:
\begin{eqnarray}
  \frac{\chi^{\mathrm {disc}}_{\mathrm {RMT}}}{N\Sigma^2} &=&
   1-K_0(2u)I_0(2u)+\Bigl[ K_0(2u)-2uK_1(2u) \Bigr] F(2u)
      \nonumber \\
  &&{}-\biggl\{2uK_0(2u) F(2u)
  -2u\Bigl[K_0(2u)I_0(2u)+K_1(2u)I_1(2u)\Bigr]\biggr\}^2 \;.
\end{eqnarray}

In Ref.~\cite{Ber2} it was demonstrated for $\chi^{\mathrm {disc}}$
that chRMT describes the Monte Carlo data perfectly up to values of
$m$ which scale like $1/L^2$ (the analogue of the Thouless energy).
This is also true for $\sigma$ and $\chi^{\mathrm {conn}}$.

We want to understand the behavior of the data beyond the Thouless
energy using (quenched) chPT. Chiral perturbation theory uses
effective actions for the Goldstone bosons to describe the mass and
volume dependence of, e.g., the chiral condensate and the
susceptibilities.  The effective action depends only on the symmetries
and the corresponding breaking pattern.  The predictions of chRMT are
equivalent \cite{chiralpert} to the leading-order results from the
so-called $\epsilon$-expansion in chPT, where $m = O(\epsilon^4)$ and
$1/L = O(\epsilon)$. The coefficients in this expansion are nontrivial
functions of $u = m \Sigma L^4$.  On the other hand, standard chiral
perturbation theory (the so-called $p$-expansion) sets $m = O(p^2)$,
$1/L = O(p)$ and expands in powers of $p$ (see, e.g., \cite{galeu}).

Applying chPT in the present context we have to deal with two
technical problems.  Since the data that we want to analyze are taken
in the quenched approximation we ought to work with the quenched
version of chiral perturbation theory. Secondly, we work at rather
strong coupling where we cannot expect the continuum symmetries to be
already effectively restored. So we should consider only symmetries
which are exact on the lattice (and the corresponding Goldstone
bosons), a situation usually not dealt with in the literature.

Below the Thouless energy the first problem becomes trivial when we
use the chRMT results, because these depend explicitly on the number
of flavors which we can set equal to zero. Furthermore they also
depend on the topological charge $\nu$ (i.e.\ the number of zero modes
of the Dirac operator), which suggests a solution to the second
problem: At strong coupling, the Dirac operator of staggered fermions
has no exact topological zero modes due to lattice artifacts, hence
the lattice results should be compared with the case $\nu = 0$
\cite{r5}.  Indeed, one finds very good agreement.

Above the Thouless energy, in the regime of the $p$-expansion, we
should use (partially) quenched chPT taking into account the pattern
according to which chiral symmetry is spontaneously broken in the case
of staggered fermions with gauge group SU(2).  To the best of our
knowledge, an analysis of this particular case has not been done
before. In the following, we present our own, somewhat unorthodox,
approach to this problem in order to avoid the intricacies of quenched
chiral perturbation theory.

Starting point is a partition function of the form:
\begin{equation} \begin{array}{l} \displaystyle
  Z(m_v, m_s, L)
\\ \displaystyle
\propto \exp \left[ V T(m_v , m_s)
 -\frac{1}{2} \sum_Q \left\{ K_Q
 \sum_p \ln \left(\ph^2 + m_Q^2(m_v, m_s) \right) \right\} \right] .
 \end{array}
 \label{Z}
\end{equation}
$T(m_v, m_s)$ is the classical (or saddle-point) contribution to the
free energy, which we will assume is a smooth function of the quark
masses, and independent of lattice volume.  The double sum represents
the 1-loop contribution coming from light composite bosons. The sum
runs over the allowed momenta $p$ ($p_\mu = 2 \pi n_\mu /L$ with
integer $n_\mu$) and over particle type $Q$. $K_Q$ is the multiplicity
of the particles of type $Q$. On the lattice we will use the
expression
\begin{equation}
 {\hat p}^2 \equiv 2 \sum_\mu (1 - \cos p_\mu)
\end{equation}
for the boson kinetic term, where $p_\mu$ is expressed in lattice
units.

We have introduced two quark masses, valence quarks with mass $m_v$,
and sea quarks with mass $m_s$. We have $N_v$ `generations' of the
quarks with mass $m_v$, and $N_s$ generations of the quarks with mass
$m_s$. In the continuum limit, each staggered `generation' yields four
fermion flavors. We will take the limit $N_v \rightarrow 0$, so that
the valence quarks are quenched.

To use Eq.~(\ref{Z}) we need the masses and multiplicities
of all the light bosons. For mesons made of two different
quark flavors ${\bar q}_i  q_j$ we use the expression
\begin{equation}
 m^2 = A (m_i + m_j)/2 .
\end{equation}
This applies to $(N_s + N_v)(N_s+N_v-1)$ bound states.

\begin{figure}[htb]
 \begin{center}
 \epsfig{file=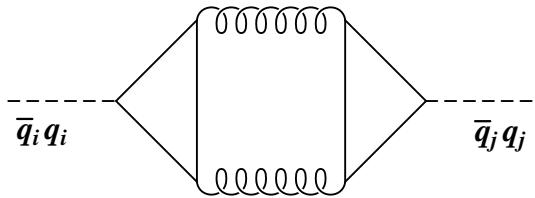,width=7cm}
 \end{center}
 \caption{ Annihilation diagram for pseudoscalar mesons.}
 \label{ann}
 \end{figure}

For the remaining $N_s + N_v$ `flavor-diagonal' mesons we should also
consider annihilation (see Fig.~\ref{ann}).
However, with staggered fermions there is
an anomaly-free U(1) symmetry, which would cause the
amplitude for ${\bar q}_i q_i \rightarrow {\bar q}_j q_j$
to vanish if either $m_i = 0$ or $m_j = 0$. Therefore we
expect the annihilation amplitude to be proportional
to $m_i m_j$ for small quark masses. In the final formulae this
leads to contributions which possess only a mild $m$--dependence
(there is at least one more power of $m$ than in the leading
contribution).  
Hence these are hard to distinguish from the smooth background and
cannot be fitted reliably. To leading order, it is consistent to
neglect the annihilation terms, and we end up with $N_v$ ($N_s$)
states with $m^2 = A m_v$ ($m^2 = A m_s$) in the `flavor-diagonal'
sector.  (Note that for the gauge group SU(3) there are special cases
where the annihilation terms contribute to leading order and can
therefore not be neglected.)

For the SU(2) gauge group the symmetry when all quarks are
massless is $\mathrm{U}(2 N_v + 2 N_s)$, which spontaneously breaks
to $\mathrm{O}(2 N_v + 2 N_s)$ \cite{klu}. This is further broken to
$\mathrm{O}(2 N_v) \times \mathrm{O}(2 N_s)$ if the valence and sea
quarks are given different bare masses. The extra symmetry present
when the gauge group is SU(2) transforms mesons into
baryons ($q_i q_j$ and ${\bar q_i} {\bar q_j}$
states), which thus have the same mass as the mesons.
These states will have mass-squared given by the same formula
as the different-flavor mesons, i.e., $m^2 = A (m_i + m_j)/2$.
Therefore we have to include more light bosons in our partition
function leading to the spectrum given in Table \ref{su2spect}.

\renewcommand{\arraystretch}{1.5}
\begin{table}
\caption{ The light particle spectrum for gauge group SU(2).}
\label{su2spect}
\begin{center}
 \begin{tabular}{|c||c|}
 \hline
  $m^2$ & multiplicity \\ \hline
 $A m_v$  & $2 N_v^2 + N_v$   \\
 $A m_s$  & $2 N_s^2 + N_s$   \\
 $A (m_v + m_s)/2 $&$ 4 N_v N_s $ \\
  \hline
 \end{tabular}
\end{center}
\end{table}

In SU(2) we just sum over the non-degenerate eigenvalues
(according to the RMT conventions), so our definition of the
chiral condensate $\sigma$ is
\begin{equation}
 \sigma(m_v, m_s, L) = \lim_{N_v \rightarrow 0}
 \left( \,\frac{1}{V} \frac{1}{2 N_v} \,\frac{1}{Z}
 \,\frac{\partial Z}{\partial m_v} \right) \,.
 \label{sigdef2}
\end{equation}
If we substitute the multiplicities of Table~\ref{su2spect} into
Eq.~(\ref{Z}) we get
\begin{eqnarray}
 \sigma(m_v, m_s ,L)  & = &  S + C_c m_v + C_d 2 N_s m_s
 - \frac{A}{4 L^4} \sum_p \frac{1}{\ph^2 + A m_v}
 \nonumber \\
 & & {} - \frac{A N_s }{L^4}
 \sum_p \frac{1}{2 \ph^2 + A (m_v + m_s)} \,,
 \label{su2sig}
\end{eqnarray}
where the first three terms parametrize the classical background
contribution.  Differentiating $\sigma$ to give the susceptibilities
we find for identical valence and sea quark masses, $m_v = m_s = m$,
\begin{eqnarray}
 \chi^{\mathrm {conn}}(m, m, L) & = & C_c
 + \frac{ A^2 (1 + N_s)}{4 L^4} \sum_p \frac{1}{(\ph^2 + A m)^2} \,,
 \label{chpt.suscon} \\
 \chi^{\mathrm {disc}}(m, m, L) & = & C_d
 + \frac{ A^2 }{8 L^4} \sum_p \frac{1}{(\ph^2 + A m)^2} \,.
 \label{chpt.susdis}
\end{eqnarray}
Sending finally also $N_s$ to zero we arrive at formulae appropriate
for comparison with our quenched Monte Carlo data. Note that the
leading behavior of these chPT formulae for $m \to 0$ coincides with
the leading term for $m \to \infty$ of the chRMT formulae.  This is in
accordance with the existence of a mass range where both theories
apply.

In the thermodynamic limit we obtain from (\ref{su2sig}),
(\ref{chpt.suscon}), and (\ref{chpt.susdis}) chiral logarithms:
\begin{eqnarray}
 \sigma (m) & = & \left( S - \frac{1}{4} A a_0 \right)
    - \frac{A^2}{64 \pi^2} m \ln (Am)
    + \left( C_c - \frac{1}{4} A^2 a_1 \right) m
  \nonumber \\
  & &  {} + \frac{A^3}{512 \pi^2} m^2 \ln (Am)
   - \frac{1}{4} A^3 a_2 m^2 + \cdots \,,
\end{eqnarray}
\begin{eqnarray}
 \chi^{\mathrm {conn}}(m) & = &  - \frac{A^2}{64 \pi^2} \ln (Am)
      + \left( C_c - \frac{1}{4} A^2 a_1
                - \frac{A^2}{64 \pi^2} \right)
  \nonumber \\
  & &  {} + \frac{A^3}{256 \pi^2} m \ln (Am)
  + A^3 \left( \frac{1}{512 \pi^2} - \frac{1}{2} a_2
          \right) m + \cdots \,,
\end{eqnarray}
\begin{eqnarray}
 \chi^{\mathrm {disc}}(m) & = &  - \frac{A^2}{128 \pi^2} \ln (Am)
      + \left( C_d - \frac{1}{8} A^2 a_1
      - \frac{A^2}{128 \pi^2} \right)
  \nonumber \\
  & &  {} + \frac{A^3}{512 \pi^2} m \ln (Am)
       + A^3 \left( \frac{1}{1024 \pi^2}
              - \frac{1}{4} a_2 \right) m + \cdots
\end{eqnarray}
for $N_s = 0$ and $m_v = m_s = m$. The numerical constants $a_0$,
$a_1$, and $a_2$ take the values $a_0 = 0.1549$, $a_1 = -0.03035$,
$a_2 = 0.002776$.

To highlight the special properties of the SU(2) gauge group with
staggered fermions, we will now compare with the conventional
continuum calculation with gauge group SU(3).  In this case there are
no light `Goldstone baryons', so we only need to consider meson
states. The other important difference is that in the continuum the
chiral U(1) has an anomaly. This means that the annihilation term in
the meson mass matrix does not need to vanish in the chiral limit.
This leads us to the following ansatz for the mass-squared matrix,
$M^c$, for the flavor-diagonal mesons:
\begin{equation}
 M^c = A \, \mathrm{diag} \left( m_v, \ldots, m_v, m_s,
                                \ldots, m_s \right)
  + m_0^2 \left( \matrix{
  1 & \ldots & 1 \cr
                 \vdots & \ddots& \vdots \cr
  1 & \ldots & 1
 } \right)
 \label{contmassmat}
\end{equation}
with $N_v$ ($N_s$) entries $Am_v$ ($Am_s$) on the diagonal of the
first part. Here $m_0$ is a constant with the dimensions of mass,
which does not vanish in the chiral limit.

Besides $N_v -1$ eigenvalues equal to $A m_v$ and $N_s -1$ eigenvalues
equal to $A m_s$ this mass matrix has two non-trivial eigenvalues.
The important thing to note is that one of them does not vanish in the
chiral limit but goes to the value $(N_v + N_s ) m_0^2$.  This is the
state corresponding to the continuum $\eta^\prime$.

Using the continuum multiplicities and masses in our formula for
$Z$ and omitting the smooth background for simplicity leads to
\begin{equation}
 \sigma^c(m,m,L) = \frac{A}{2 L^4} \sum_p
 \frac{m_0^2 (1 - N_s^2) - N_s (\ph^2 + A m)}
 {(\ph^2 + A m)(\ph^2 + N_s m_0^2 + A m) }
 \label{sigc}
\end{equation}
(the case of equal valence and sea quark masses is sufficient to
illustrate the main differences between staggered and continuum
fermions). For $N_s \neq 0$ the $L \rightarrow \infty$
limit of this $\sigma$ is
\begin{eqnarray}
  \sigma^c(m,m,\infty) &=&
 \frac{A}{2 N_s}
 \frac{A m}{16 \pi^2} \ln \left[ \frac{A m}{N_s m_0^2 + A m} \right]
 \nonumber \\ && {}
 - \, \frac{A N_s}{2} \left( a_0 + \frac{A m}{16 \pi^2} \ln (A m)
 + a_1 A m \right)
 \nonumber \\ && {}
 - \, \frac{A}{2} \left( \frac{m_0^2 }{16 \pi^2}
         \ln (N_s m_0^2 + A m)
 + a_1 m_0^2 \right) + \cdots \, .
\end{eqnarray}
At small $m$ the most severe singularity is of the form $m \ln m$.

However in the quenched limit the behaviour is different.
Then Eq.~(\ref{sigc}) reduces to
\begin{equation}
 \sigma^c(m,m,L) = \frac{A}{2 L^4} \sum_p
 \frac{m_0^2 }
 {(\ph^2 + A m)^2} \,,
\end{equation}
which is more singular at small $m$. The thermodynamic limit
of this expression is
\begin{equation}
  \sigma^c(m,m,\infty) = \frac{ A m_0^2}{2} \left[
 - \frac{1}{16 \pi^2} \left( \ln (A m) + 1 \right) - a_1
 + \cdots \right] \,.
\end{equation}
Now we have a more severe singularity ($\propto \ln m$) in the small
mass limit.  This is the well-known quenched chiral logarithm.  We do
not get this logarithm in our regime because the relevant U(1)
symmetry is not anomalous.

\begin{figure}
\epsfig{file=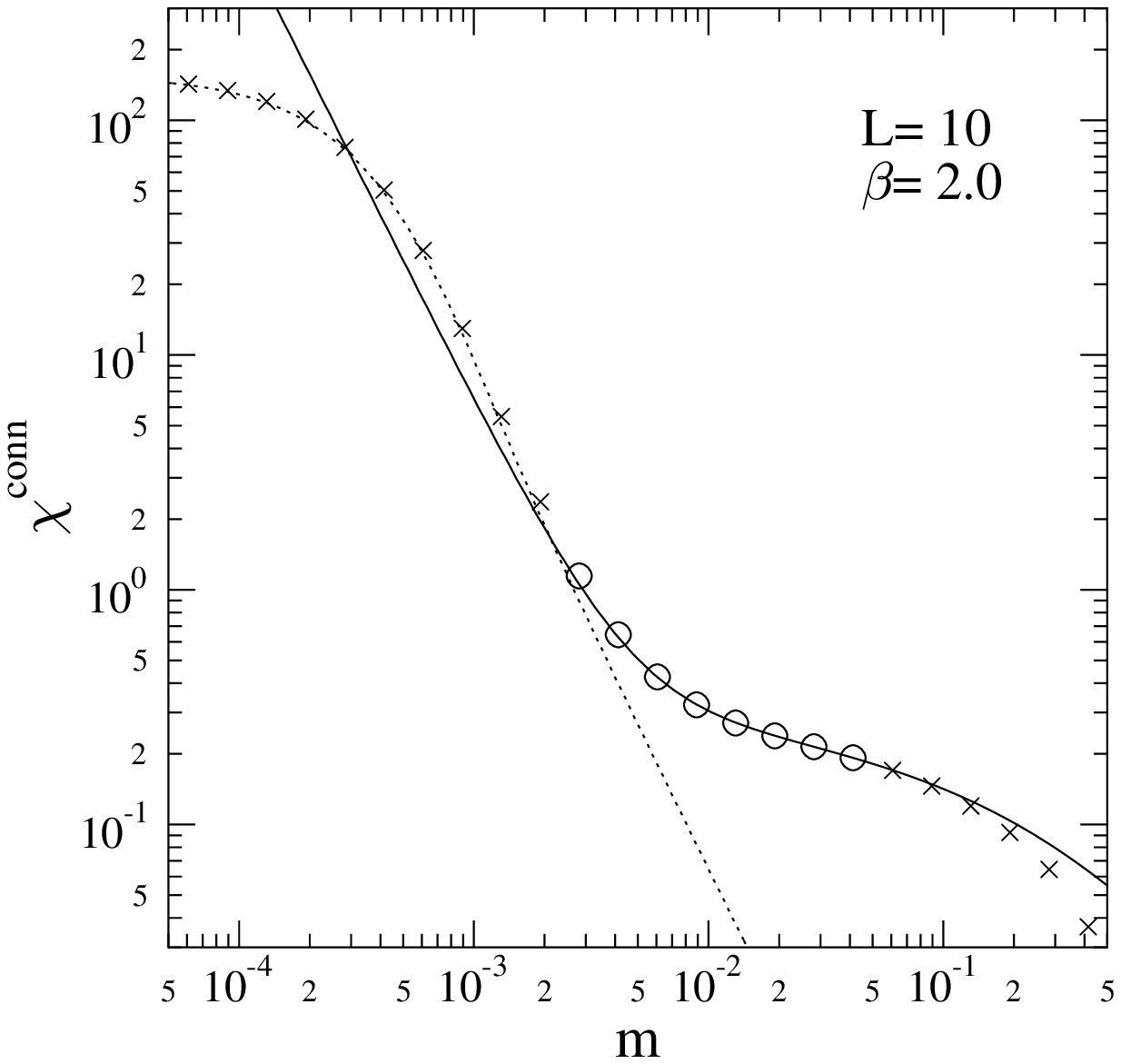,width=7cm}
\hspace{0.5cm}
\epsfig{file=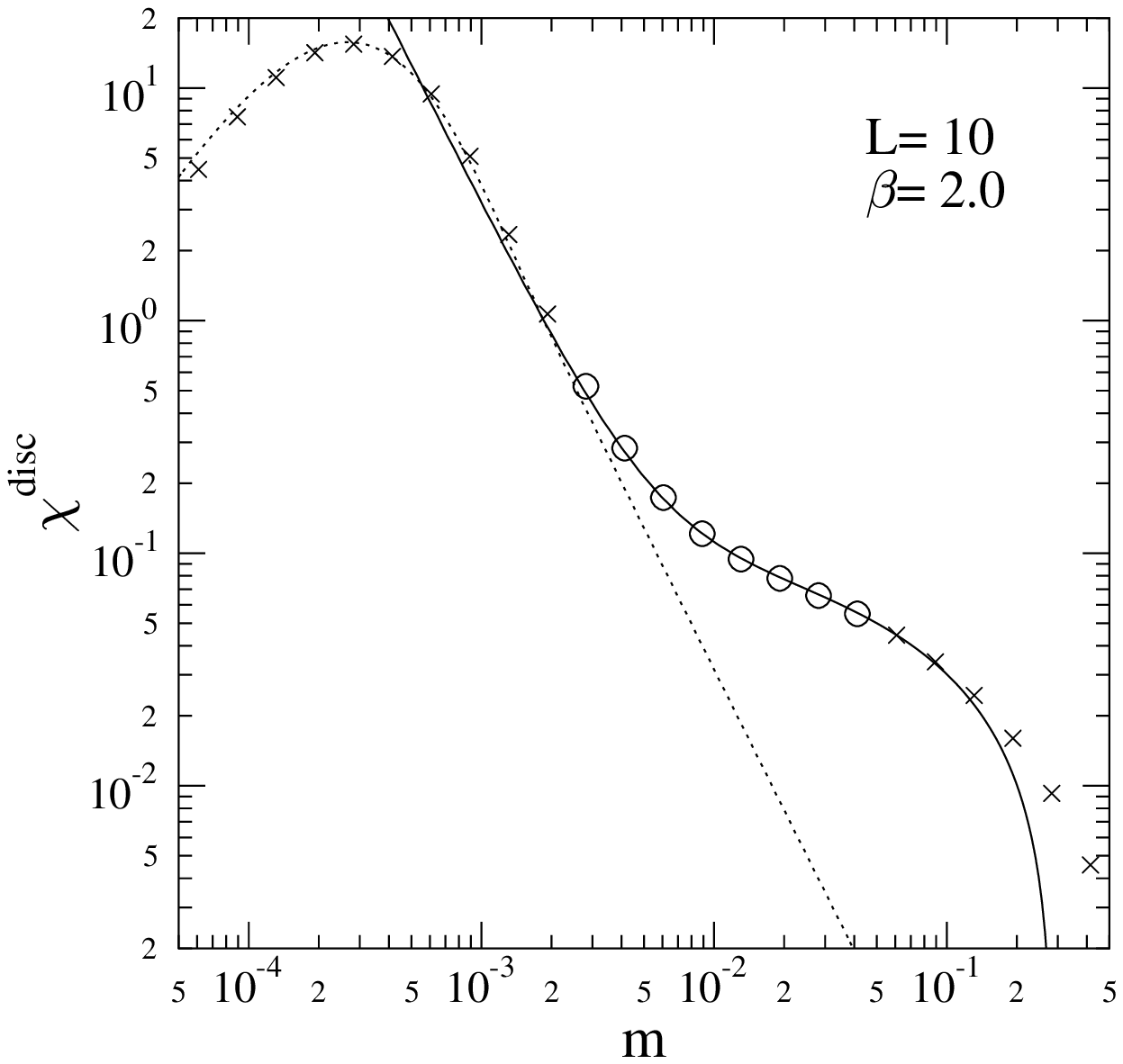,width=7cm}
\\[-0.6cm]%
\epsfig{file=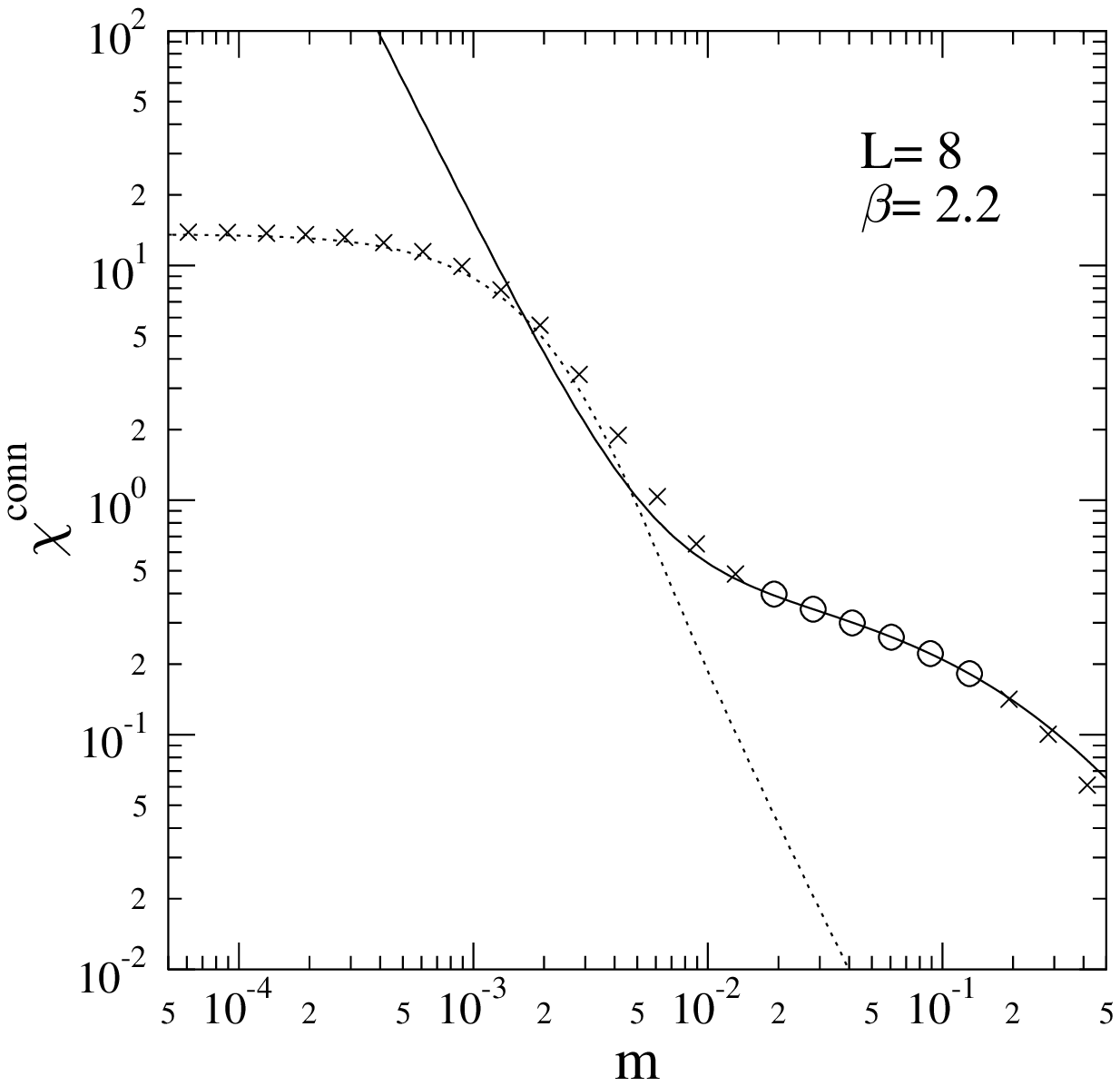,width=7cm}
\hspace{0.5cm}
\epsfig{file=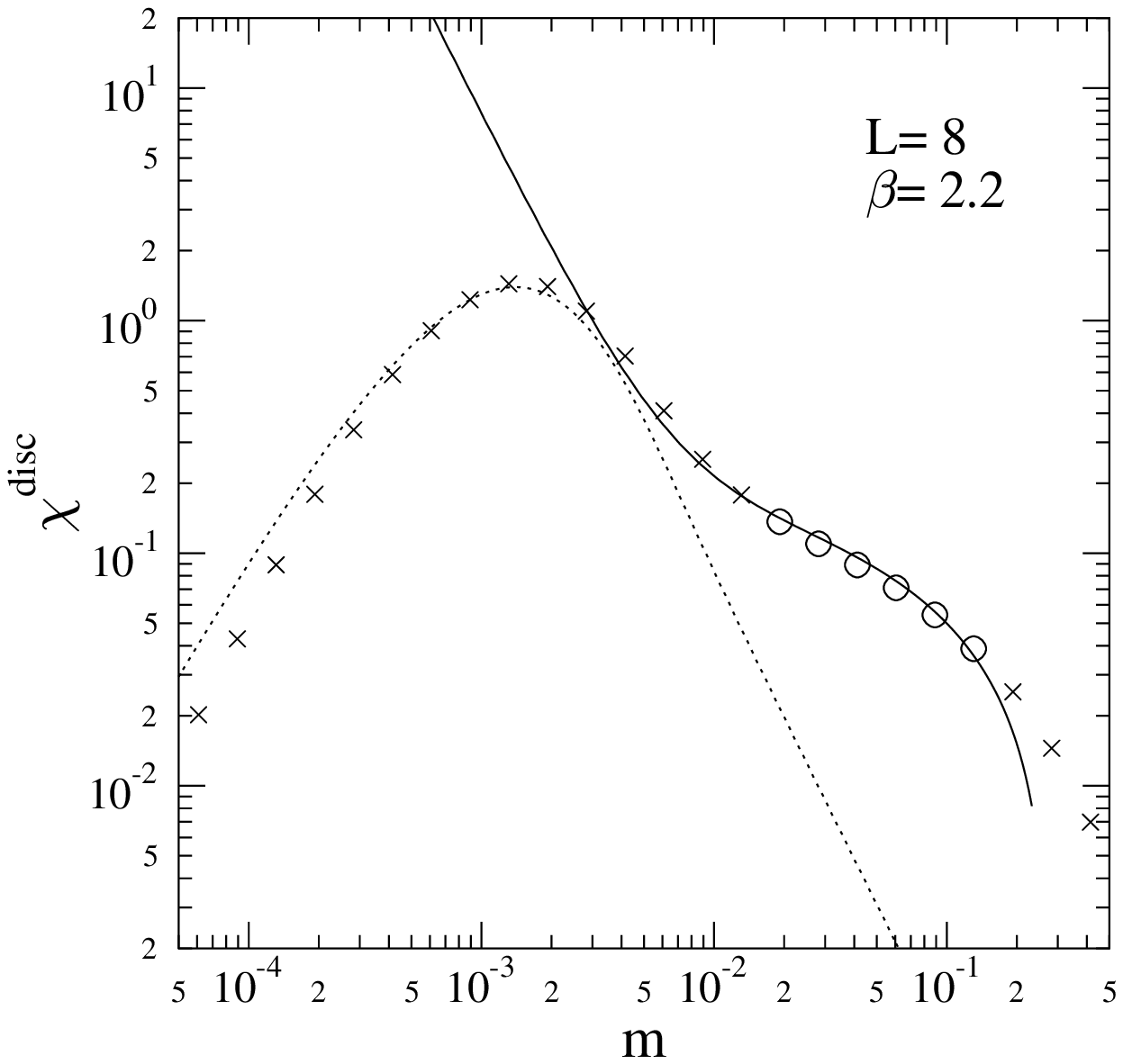,width=7cm}
\\[-0.6cm]%
\epsfig{file=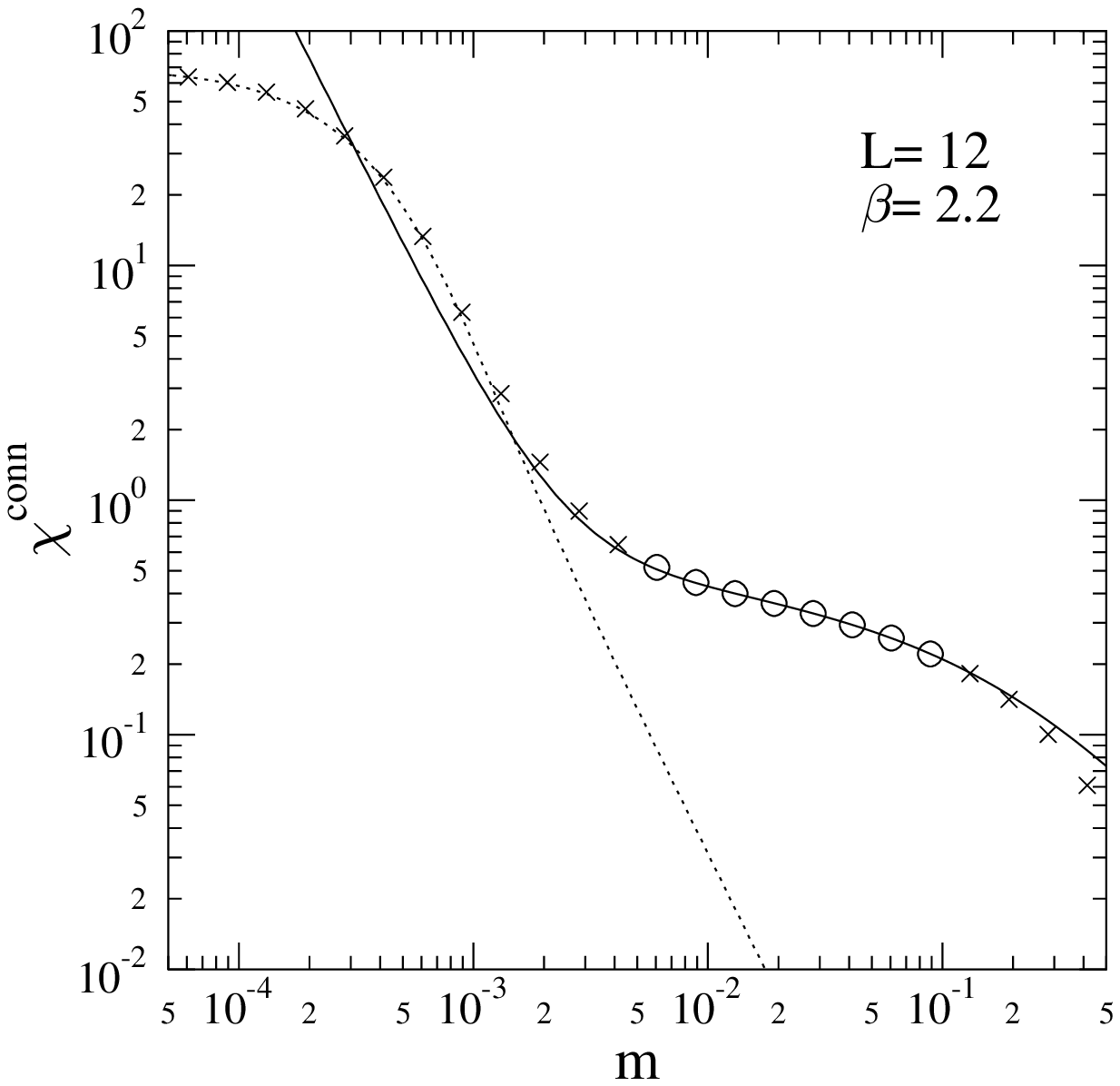,width=7cm}
\hspace{0.5cm}
\epsfig{file=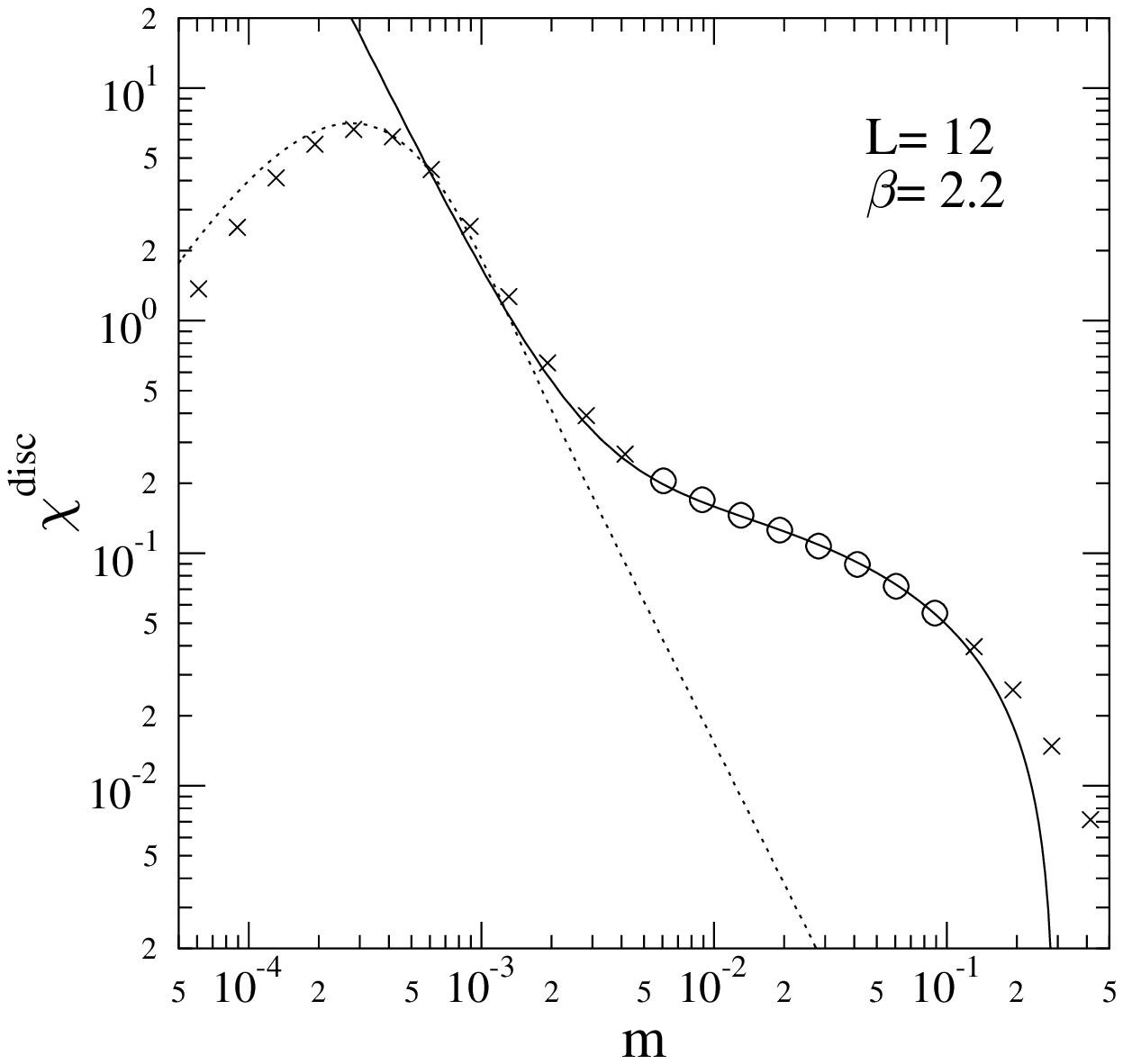,width=7cm}
\caption{Connected and disconnected susceptibility for selected values
  of $L$ and $\beta$.  The chRMT results are shown as dotted lines,
  the chPT fits are displayed as full curves. Data points within the
  fit range are marked by circles, else by crosses. For clarity the
  statistical errors of the data points are not shown, in most cases
  they are smaller than the symbols.}
\label{chiplot}
\end{figure}

We have confronted our chPT formulae with lattice Monte Carlo data.
We used staggered fermions in the quenched approximation and two
values of the coupling strength $\beta$, $\beta=2.0$ and $\beta=2.2$.
The lattice sizes and numbers of configurations are given in
Table~\ref{results}.  The chPT formulae are able to describe the
deviations from the RMT results beyond the Thouless energy very well.
Examples of joint fits of $\chi^{\mathrm {conn}}$ and $\chi^{\mathrm
  {disc}}$ with (\ref{chpt.suscon}) and (\ref{chpt.susdis}),
respectively, are presented in Fig.~\ref{chiplot}.  It seems that for
the connected susceptibility our lattices are not large enough to
exhibit clearly a domain where both chRMT and chPT (to the order
considered) are applicable, although there is a trend towards the
formation of such a window with increasing $L$.

Joint fits of $\chi^{\mathrm {conn}}$ and $\chi^{\mathrm {disc}}$ lead
to the values for $A$ given in Table~\ref{results}.  The error has
been estimated from the variation of the results under changes of the
fit interval. This is a somewhat subjective procedure of determining
the error and goes beyond the purely statistical contribution. However
we consider it to be more realistic than using the MINUIT errors. The
smaller errors on the larger lattices thus reflect the increasing
reliability of our numbers.

Due to the Gell-Mann--Oakes--Renner relation the parameter $A$ is
related to the chiral condensate $\Sigma$ (for infinite volume and
vanishing mass) and the pion decay constant $f_\pi$ by
\begin{equation}
 A = \frac{2 \Sigma}{f_\pi^2}  \;.
\end{equation}
The chiral condensate $\Sigma$ can be determined independently,
e.g.\ from the mean value
of the smallest eigenvalue according to the RMT formula \cite{r5}
\begin{equation}
 \Sigma = \frac{2.06636\cdots}{L^4
             \langle \lambda_{\mathrm {min}} \rangle } \;.
\end{equation}
The corresponding numbers together with the resulting values for
$f_\pi$ are also given in Table~\ref{results}.  Note that except for
the smallest lattices the results are nearly independent of $L$
demonstrating the reliability of our procedure.
\begin{table}
\caption{Simulation parameters together with fit results for the
  chPT parameter $A$ and the chiral condensate $\Sigma$. The value of
  the pion decay constant $f_\pi$ following from the
  Gell-Mann--Oakes--Renner relation is given in the last column.}
\label{results}
\vspace{1.0cm}
\begin{center}
\begin{tabular}{crr@{\hspace{1.5cm}}lll}
\hline
$\beta$ & \multicolumn{1}{c}{$L$} & configs & \multicolumn{1}{c}{$A$}
      & \multicolumn{1}{c}{$\Sigma$} & \multicolumn{1}{c}{$f_\pi$}\\
\hline
 2.0 & 4  & 49978 & 13.5(9) & 0.1157(2)  & 0.131(4) \\
 {}  & 6  & 24976 & 11.6(2) & 0.1258(3)  & 0.147(1) \\
 {}  & 8  & 14290 & 10.9(1) & 0.1253(4)  & 0.1516(7)\\
 {}  & 10 &  4404 & 10.4(1) & 0.1239(7)  & 0.1544(9)\\
\hline
 2.2 & 6  & 22288 & 14.5(5) & 0.0540(2)  & 0.086(2) \\
 {}  & 8  & 13975 & 13.8(2) & 0.0568(2)  & 0.0907(7)\\
 {}  & 10 &  2950 & 13.6(1) & 0.0573(4)  & 0.0918(5)\\
 {}  & 12 &  1382 &13.4(1) & 0.0576(6)  & 0.0927(6)
\end{tabular}
\end{center}
\vspace{1.0cm}
\end{table}

In conclusion, we have verified predictions of chiral random matrix
theory and (quenched) chiral perturbation theory by means of lattice
Monte Carlo data.  In particular, we have investigated finite-volume
effects and found a domain of common validity of both theories.  It
was crucial to base the analysis on the lattice symmetries and not on
the continuum symmetries.  We were also able to determine the pion
decay constant. The extension of the present analysis to the case of
gauge group SU(3) will be the subject of an upcoming publication.

\begin{ack}
  It is a pleasure to thank I.~Zahed for an early discussion and
  M.~Golterman, N.~Kaiser, and J.J.M.~Verbaarschot for helpful
  comments. This work was supported in part by DFG. SM, AS, and TW
  thank the MPI f\"ur Kernphysik, Heidelberg, for hospitality and
  support. The numerical simulations were performed on a CRAY T90 at
  the Forschungszentrum J\"ulich, on a CRAY T3E at the HLRS Stuttgart,
  and on a Cray T90 at the Leibniz-Rechenzentrum M\"unchen.
\end{ack}

\end{document}